\documentclass[aps,prb,reprint,groupedaddress,showpacs]{revtex4-1}
\usepackage{graphicx}
\usepackage{siunitx}
\renewcommand{\vec}[1]{\mathbf{#1}}
\newcommand{\dd}{\text{d}}
\begin{document}

\title{Multiscale and multimodel simulation of Bloch point dynamics}

\author{Christian Andreas}
\email[Corresponding author: ]{christian.andreas@ipcms.unistra.fr}
\affiliation{Peter Gr\"unberg Institut (PGI-6), Forschungszentrum J\"ulich GmbH, D-52428 J\"ulich, Germany}
\affiliation{Institut de Physique et Chimie des Mat\'eriaux de Strasbourg,
Universit\'e de Strasbourg, CNRS UMR 7504, Strasbourg, France}%

\author{Attila K\'akay}
\affiliation{Peter Gr\"unberg Institut (PGI-6), Forschungszentrum J\"ulich GmbH, D-52428 J\"ulich, Germany }%

\author{Riccardo Hertel}
\affiliation{Institut de Physique et Chimie des Mat\'eriaux de Strasbourg,
Universit\'e de Strasbourg, CNRS UMR 7504, Strasbourg, France}%

\date{\today}
\begin{abstract}
We present simulation results on the structure and dynamics of micromagnetic point singularities with atomistic resolution. This is achieved by embedding an atomistic computational region into a standard micromagnetic algorithm. Several length scales are bridged by means of an adaptive mesh refinement and a seamless coupling between the continuum theory and a Heisenberg formulation for the atomistic region. The code operates on graphical processing units and is able to detect and track the position of strongly inhomogeneous magnetic regions. This enables us to reliably simulate the dynamics of Bloch points, which means that a fundamental class of micromagnetic switching processes can be analyzed with unprecedented accuracy. We test the code by comparing it with established results and present its functionality with the example of a simulated field-driven Bloch point motion in a soft-magnetic cylinder.
\end{abstract}
\maketitle

\section{\label{sec:intro}Introduction}
Textbooks on ferromagnetism typically list three fundamental modes for magnetization reversal: buckling, curling, and rotation in unison\cite{aharoni_introduction_2000,brown_micromagnetics_1963,kronmuller_general_2007}. Those basic instabilities were derived analytically for homogeneously magnetized ferromagnetic cylinders in an external magnetic field\cite{frei_critical_1957}. Over the past two decades, high-resolution imaging techniques and advanced micromagnetic simulation studies have shown that the magnetization reversal is usually a much more complex process than predicted by analytic theory. It may involve the nucleation and the propagation of vortices, antivortices, domain walls of various types, and it is usually accompanied by a broad spectrum of spin wave excitations. The reason for such complications are finite-size effects. The three fundamental reversal modes are only correct for the hypothetical case of an infinitely extended ferromagnetic cylinder; but the end of a real cylinder acts as a nucleation site, leading to a completely different type of reversal. The importance of the end of a cylinder was investigated by Arrott and coworkers \cite{arrott_point_1979,hubert_arrotts_1998}. They predicted that point singularities play a decisive role in the switching of soft-magnetic cylinders. Such point singularities of the magnetization are known as Bloch points. They were described by Feldtkeller \cite{feldtkeller_mikromagnetisch_1965} and later by D\"oring \cite{doring_point_1968} and they represent a fundamental class of micromagnetic structures.

\section{Topological defects and singularities}

The simulation of Bloch points is connected with great difficulties. A reason for this is that they represent singularities in the continuum theory of micromagnetism. As a prototype and as the simplest version of a large family of Bloch point structures, one can imagine such singularities as a centrosymmetric hedgehog-type arrangement, where the magnetization vectors are radially oriented away from the Bloch point. For symmetry reasons, no magnetization direction can be assigned to the point at the center of such a structure. This violates a fundamental aspect of micromagnetic theory, which treats the magnetization as a continuous and smooth vector field with constant magnitude at every point within the ferromagnet. Bloch points are hence topological defects in the vector field of the magnetization. Around these points the magnetization is maximally inhomogeneous, which causes the exchange energy density to diverge. In spite of this singularity in the energy density, the total energy of a Bloch point structure remains finite. The strong inhomogeneity generates another, from a practical viewpoint even more important difficulty: The exchange energy density formulation in micromagnetism relies on the assumption of a smooth variation of the magnetization on the length scale of the atomic lattice constant. Based on this assumption, which is perfectly valid in the case of ordinary micromagnetic structures like vortices or domain walls, it is easy to derive that the exchange field is proportional to the Laplacian of the magnetization, which is used, {\em e.g.}, in any micromagnetic code based on the Landau-Lifshitz-Gilbert equation \cite{landaulifshitz-exchange,gilbert_orig_abst_1955,gilbert_phenomenological_2004}. Removing the approximation of a slow spatial variation makes the entire calculation of the exchange field questionable. In addition to the difficulties concerning the theoretical foundations, the simulation of singularities in continuum theories is a delicate subject. The discretization error displays a power-law behavior as a function of the cell size \cite{zienkiewicz_finite_2005}, making it particularly difficult to obtain reliable solutions. The most careful studies on Bloch points in the literature therefore make use of various grids and extrapolate the computed values to infinite discretization density \cite{thiaville_micromagnetic_2003,gliga_energy_2011}. This procedure is tedious but it can at least remove the numerical artifacts. 

\section{The multiscale problem}

It appears that all difficulties connected with Bloch points stem from micromagnetic theory. If an atomistic Heisenberg model is used, nothing prevents neighboring atomistic magnetic moments from being strongly misaligned. Moreover, once the assumption of a continuous vector field of the magnetization is replaced by a discrete set of magnetic moments and the ferromagnetic exchange is treated with a Heisenberg term instead of the micromagnetic formulation, the singularities disappear and the problems dissolve. However, such an approach implies an atomistic treatment of the entire ferromagnet -- a model that is extremely expensive from a computation point of view and which constitutes a dramatic waste of resources since almost the entire ferromagnet could safely be simulated at significantly lower computational costs within the framework of micromagnetism. One might thus be tempted to discard the regions outside the Bloch point and investigate only the volume containing the Bloch point. However, as pointed out in the textbook of Hubert and Sch\"afer \cite{hubert_magnetic_1998} (p.~252) "a singularity has always to be analyzed together with the environment into which it is embedded". To better understand this, an analogy with basic astronomy might help. In stars, the weak but long-range gravitational force provides the confinement of a plasma with sufficiently high energy density to overcome the Coulomb barrier in nuclear fusion processes. Similarly, the relatively weak but cumulative long-range magnetostatic interaction can stabilize a highly energetic Bloch point in the volume of a sufficiently large ferromagnet. The Bloch point structure would immediately dissolve if it was removed from the mesoscopic environment into which it is embedded. The situation is therefore extremely complicated to treat numerically, since disparate different length scales and energy densities have to be combined. A recent approach to treat Bloch points consist in dropping the micromagnetic assumption of constant magnitude of the magnetization and allowing for its gradual reduction to zero in the vicinity of the Bloch point by means of the Landau-Lifshitz-Bloch equation \cite{lebecki_key_2012}. Other authors have suggested a phenomenological Landau-type energy term to account for the reduction of the magnetization in Bloch points \cite{galkina_phenomenological_1993}. This has provided the basis for a remarkable numerical study on the static three-dimensional structure of Bloch points \cite{elias_magnetization_2011}. These approaches solve the topological part of the problem because they remove the character of a point defect from these structures. But as far as the exchange representation is concerned, the difficulties remain; and only a combination of an atomistic Heisenberg model with a continuum model seems to be able to overcome this obstacle. We have developed such an appoach that we describe in detail in this article. A multiscale model extends our GPU accelerated micromagnetic code {\tt TetraMag} \cite{kakay_speedup_2010} which solves the Landau-Lifshitz Gilbert\cite{landaulifshitz-exchange,gilbert_orig_abst_1955,gilbert_phenomenological_2004} equation using a finite element / boundary element method. We implemented an atomistic Heisenberg model with realistic crystalline structure to treat the exchange interaction in quasi-classical approximation. The dynamics of the magnetic moments in the atomistic region is coupled to the motion of the magnetization into which the structure is embedded. In order to investigate the motion of the Bloch point, the atomistic part is free to move within the ferromagnet and it does so in a self-adaptive way, by recognizing the position of the Bloch point. 

After a short description of the fundamental equations we discuss the implementation of the multiscale / multimodel algorithm. In section \ref{sec:test} we demonstrate the functionality of the program on the basis of a well-defined system of a gyrating vortex, which is a complicated yet non-singular structure. This is to ascertain that the multiscale / multimodel simulation does not alter the results obtained from established micromagnetic theory. In the last part, we present an example where we employ the code to simulate the field-driven motion of a Bloch point in a soft-magnetic cylinder with atomistic resolution in the core region.

\section{\label{sec:theo} Basic Equations}
\begin{figure}
\includegraphics[width=\linewidth]{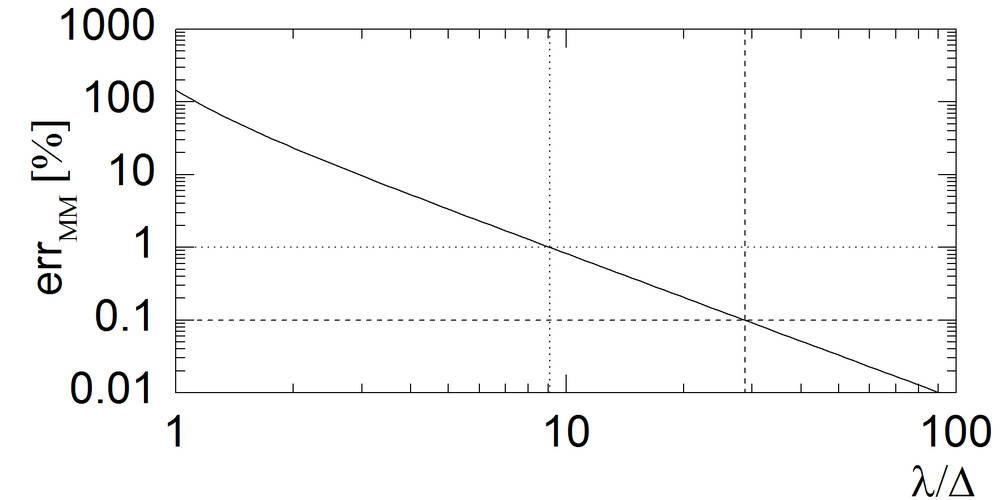}
\caption{Error estimate for the intrinsic deviation of the micromagnetic exchange energy formulation from a Heisenberg calculation. The error is displayed for the test case of a spin spiral of wave length $\lambda$, where $\Delta$ is the distance between neighboring magnetic moments along the direction of the spin spiral.\label{Fig:1a}} 
\end{figure}

For the multiscale / multimodel simulation we assume ferromagnetic crystalline samples and combine  the representation of exchange energy in the micromagnetic (MM) model with the one of the Heisenberg (Hei) formulation. In micromagnetic theory, the exchange energy density is defined as \cite{brown_micromagnetics_1978}
\begin{equation}
e_{xc} = A \sum_{\alpha}^{x,y,z} \frac{\partial \vec{m}}{\partial x_\alpha} \cdot \frac{\partial \vec{m}}{\partial x_\alpha}  
\label{Eq:excMM}
\end{equation}
where $A$ is the exchange stiffness and $\vec{m}$ the orientation of magnetization. The classical Heisenberg model represents the exchange energy density at a lattice site $i$ according to
\begin{eqnarray}
e_{xc}^{(i)} &=& -\sum_{j} \frac{J_{i,j}}{V_i}\vec{s}_i \cdot \vec{s}_j \nonumber\\
&=&  -\sum_{j} \frac{J_{i,j}}{V_i} \cos\theta_{ij} 
\label{heisenberg-exchange-energy-density}
\end{eqnarray}
where $V_i$ is the volume ascribed to the lattice site $i$, $J_{i,j}$ is the exchange constant between the magnetic moments $i$ and $j$ with orientation $\vec{s}_i$ and $\vec{s}_j$ enclosing the angle $\theta_{ij}$.

Equation (\ref{heisenberg-exchange-energy-density}) can be expanded into a Taylor series 
\begin{equation}\label{taylor}
e_{xc}^{(i)} =  -e_0 + \sum_{j} J_{i,j}/V_i \left[\frac{\theta_{ij}^2}{2} - \mathcal{O}(\theta_{ij}^4)\right] \label{Eq:excHei}
\end{equation}
with a constant offset energy density $e_0$ that can be omitted. To derive the exchange stiffness $A_i$ for a given set of exchange constants $J_{ij}$ in the vicinity of the lattice site $i$, a spin spiral with arbitrary wave length $\lambda\gg a$ can be used as a reference structure ($a$ is the lattice constant). Owing to the isotropy of Eq. (\ref{Eq:excMM}) and (\ref{heisenberg-exchange-energy-density}), the direction of the spin spiral can be chosen w.l.o.g. along the $x$ axis.
\begin{equation} 
 \vec{m}(x) = \left(
\begin{array}{c}
 \sin(\pi x / \lambda) \\
 \cos(\pi x / \lambda) \\
0
\end{array}
 \right)\quad.
\end{equation}
The values can then be calibrated such that the micromagnetic and Heisenberg exchange are identical.
By considering only the first non-vanishing part of the Taylor series, a comparison between Eq. (\ref{Eq:excMM}) and Eq. (\ref{Eq:excHei}) yields the following condition for the spin spiral:
\begin{equation} 
 \sum_{j} \frac{J_{i,j}}{2V_i}   \left(\pi \frac{\Delta x_{ij}}{\lambda}\right)^2  = A_i \frac{\pi^2}{\lambda^2} \quad,
\end{equation}
where $\Delta x_{ij}$ represents the distance between the lattice sites $i$ and $j$ along $x$. The condition $\lambda^2\neq0$ yields a direct connection between the micromagnetic exchange constant $A_i$ and the Heisenberg exchange tensor $J_{ij}$. Since the micromagnetic model acts in volumes much larger than the volume $V_i$ associated to a single atom, a straightforward way to determine the micromagnetic exchange stiffness $A$ is to compute a volume-weighted average of the values $A_i$. In a single-phase monocrystalline material such an averaging is not necessary, but it can be important in complex materials.
\begin{equation}
 A = \frac{1}{\sum\limits_i V_i} \sum_i V_i \left( \sum_j \frac{J_{ij} \Delta x_{ij}^2}{2 V_i} \right)
\label{Eq:ACalibration}
\end{equation}

For monoatomic lattices this equation allows to estimate the difference between the micromagnetic and the Heisenberg model for short wave lengths $\lambda$. On a distance $\Delta$ along the direction of the spin spiral with wave length $\lambda$, the error of the micromagnetic term, which is introduced by truncating the Taylor series (\ref{taylor}), is:
\begin{equation}
  \text{err}_{\rm MM} (\Delta,\lambda)= \frac{\pi^2 \Delta^2}{2\lambda^2}\cdot \left[1-\cos\left( \frac{\pi\Delta}{\lambda}\right)\right]^{-1}-1
\label{Eq:errMM}
\end{equation}
Fig.~\ref{Fig:1a} displays the systematic error that according to Eq.~\ref{Eq:errMM} is connected with the small-angle approximation used in the micromagnetic formulation of the exchange energy density. This equation was derived assuming a Heisenberg model with nearest-neighbor interaction. The dotted and the dashed line indicate the minimum spin wave half-length below which the micromagnetic error exceeds $0.1\%$ ($\lambda < 29 \Delta$) and $1\%$  ($\lambda < 9 \Delta$), respectively.

\section{\label{sec:impl}Implementation}
\begin{figure}
 \includegraphics[width=\linewidth]{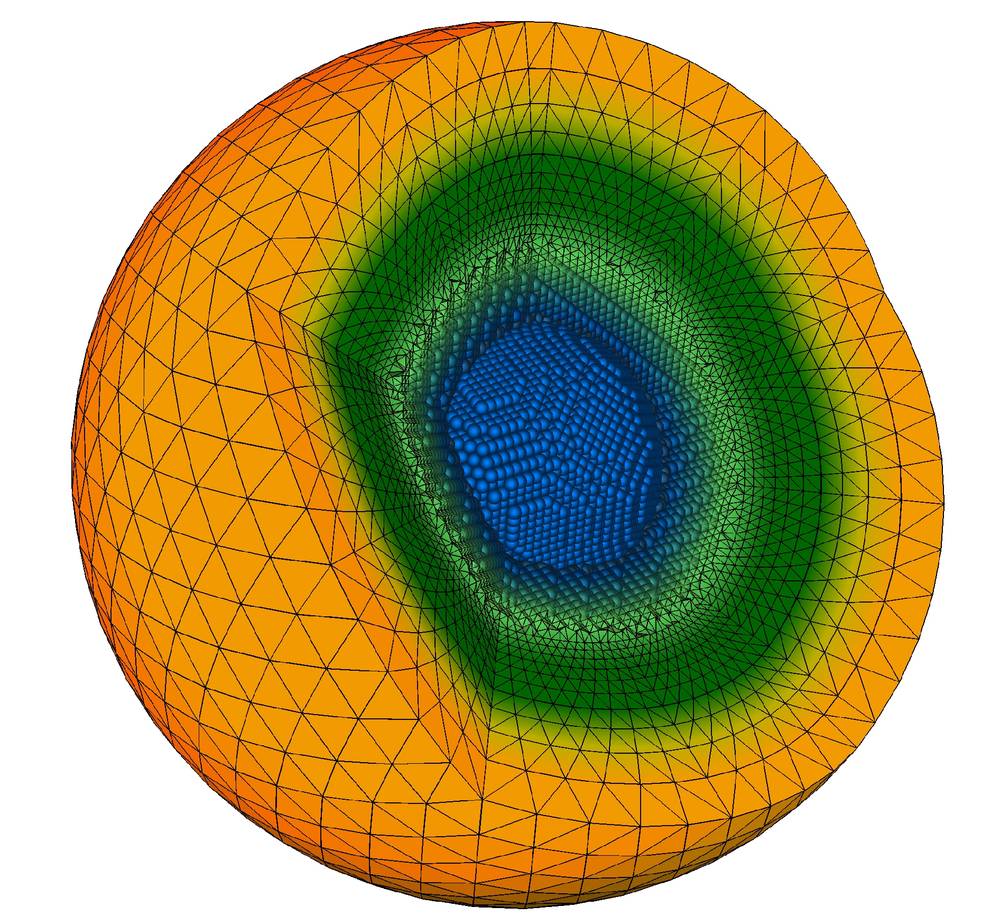}
\caption{Two different models are used to calculate the exchange interaction in the MS: Heisenberg (blue) and micromagnatic (green and orange). The green to orange shading illustrates a transition region from the outer sample to the MS, inside of which the MS stamps its properties onto the sample (green) and {\em vice versa} (orange).\label{Fig:implantedSphere}}
\end{figure}
Implementations of atomistic / continuum multiscale models in other domains of physics \cite{miller_unified_2009} are based on the refinement of the discretization grid used in the continuum model, such that the cell size is in the order of atomic distances in the region of interest. Finally, an atomistic treatment is implanted on the discretization points of the grid, to combine the equations governing the atomistic effect with those describing the continuum. As long as the region of interest remains at the same position, the interaction matrices can be preprocessed, meaning that the simulation can be performed efficiently. In the cases that we wish to study, however, the region of interest ({\em i.e.}, the Bloch point) propagates through the sample. In this case the standard procedure would require a dynamic remeshing, resulting in prohibitively high numerical costs. We therefore follow another approach: leaving the sample mesh unchanged and tracing the region of interest with a spherical structure implanted into the sample. This {\em multiscale sphere} (MS) can be preprocessed and incorporates the Heisenberg  model as well as the mesh that coarsens with increasing distance from the center of the MS. The MS used in the simulation is shown in Fig.~\ref{Fig:implantedSphere}, where a part has been removed to display the internal structure. The Heisenberg region in its center 
(blue part in Fig. \ref{Fig:implantedSphere}) represents the atomic lattice, which is surrounded by a finite element mesh with as a shell (green and orange regions in Fig. \ref{Fig:implantedSphere}) in which a gradual transition between atomistic and continuum representation is achieved. An interface region (blue to green shading in Fig. \ref{Fig:implantedSphere}) accounts for this transition from the micromagnetic to the Heisenberg  model, while the outer part is treated entirely in the framework of micromagnetism and connects the finite element calculations in the sphere with those in the surrounding volume. 

\subsection{Multiscale sphere construction}
In the core region of the sphere, a ferromagnetic crystalline structure is modeled by placing magnetic moments at the atomistic positions of the Bravais lattice. This is performed within a sphere of radius $R_1$. The atomistic positions also serve as the vertices for tetrahedral meshing in the inner part of the sphere. We thereby obtain a structure where the positions of atomic magnetic moments and vertices of finite-element cells coincide. 

The MS consists of three parts which are arranged according to the following conventions:

\begin{enumerate}
 \item In the inner core a pure Heisenberg model is employed for radii smaller then $R_H$. The location of the sphere is determined from the error estimate according to Eq.~(\ref{Eq:errMM}). We assume that a local error between $0.1 \%$ to $1 \%$ is acceptable; if this threshold is trespassed, the MS is either displaced accordingly, or a new MS is generated. 
 \item Within a transition region of thickness $\delta_T$ the exchange energy is calculated by a weighted average of the result from both, the Heisenberg model and the continuum representation. If $r$ is the distance from the center of the MS, we use 
\begin{eqnarray}
 E_{\text{multiscale}} &=& f(r)  E_{\text{Hei}} + \left[1-f(r)\right]  E_{\text{MM}} \label{Eq:transition_hei_fem}\\
& &\text{where $f(r)$ is}: \nonumber \\
f(r) &=& \frac{1}{2}\left[\cos\left(\pi \frac{r-R_1}{\delta_T}\right)+1\right]
\end{eqnarray}
\item To provide a full set of exchange partners for all magnetic moments with  $r < R_1 + \delta_T$ ghost magnetic moments are introduced in the transition region. The thickness of this shell is given by 
\begin{equation}
\delta_G = \max \left\{\left|\vec{r_i} - \vec{r_j}\right|\right\} \text{ with } J_{i,j} \neq 0 
\end{equation}
where $\vec{r_i}$ is the position of the magnetic moment $i$.
\end{enumerate}

In the shell region the atomistic core is surrounded by and attached to finite element sphere shells with radially increasing cell size. For the construction of this shell the starting point is an icosahedron as scaffolding structure, which is used to construct geodesic domes \cite{fullerpatent} over the triangular faces of the icosahedron. We choose the initial construction frequency $f_{\rm geo}^{(1)}$ (the number of subdivisions of the spherical triangle edge) so that the edge length of subtriangles matches the edge length of tetrahedra in the core of the sphere. In each further shell $n$ we reduce the frequency to $f_{\rm geo}^{(n)} = f_{\rm geo}^{(n-1)} - m$, where $m$ is a small integer; typically between 2 and 4. Smaller values of $m$ lead to a slower increase of cell size, a bigger total sphere radius and consequently more vertices. This increase of numerical costs results in smaller inhomogeneities of the mesh and therefore in a higher accuracy. The circumference radius $r_{\rm geo}^{(n)}$ of the scaffolding icosahedron for every additional shell $n$ is chosen to be $r_{\rm geo}^{(n)} = r_{\rm geo}^{(n-1)} + l_c^{(n)}$ with $l_c^{(n)}$ the average edge length of surface triangles in the new shell. Using this procedure, spherical shells are added until $l_c^{(n)}$ is larger than the average edge length used for the sample mesh. Finally, the volume between the spherical surfaces is discretized into tetrahedral simplex elements with a Delaunay algorithm using {\tt GMSH} \cite{geuzaine_gmsh_2009}. Since this triangulation is part of the preprocessing, the choice of parameters can be tuned carefully to ensure that the transition of cell sizes is smooth, thereby significantly improving the accuracy of the calculation of the exchange field and energy. 

Figure \ref{Fig:implantedSphere} shows that the entire mesh is highly regular except for the transition between the core and the shell; an irregularity which cannot be avoided since it connects the atomistic core based on a Bravais lattice to the mesoscopic spherical shell. 

At radii $r > R_H + \delta_T$ the simulations are purely micromagnetic. The radially increasing cell size of the shell region prevents an abrupt change of length scales, which could give rise to significant artifacts \cite{miller_unified_2009}.

\subsection{Multiscale regions near the surface}
\begin{figure}
 \includegraphics[width=\linewidth]{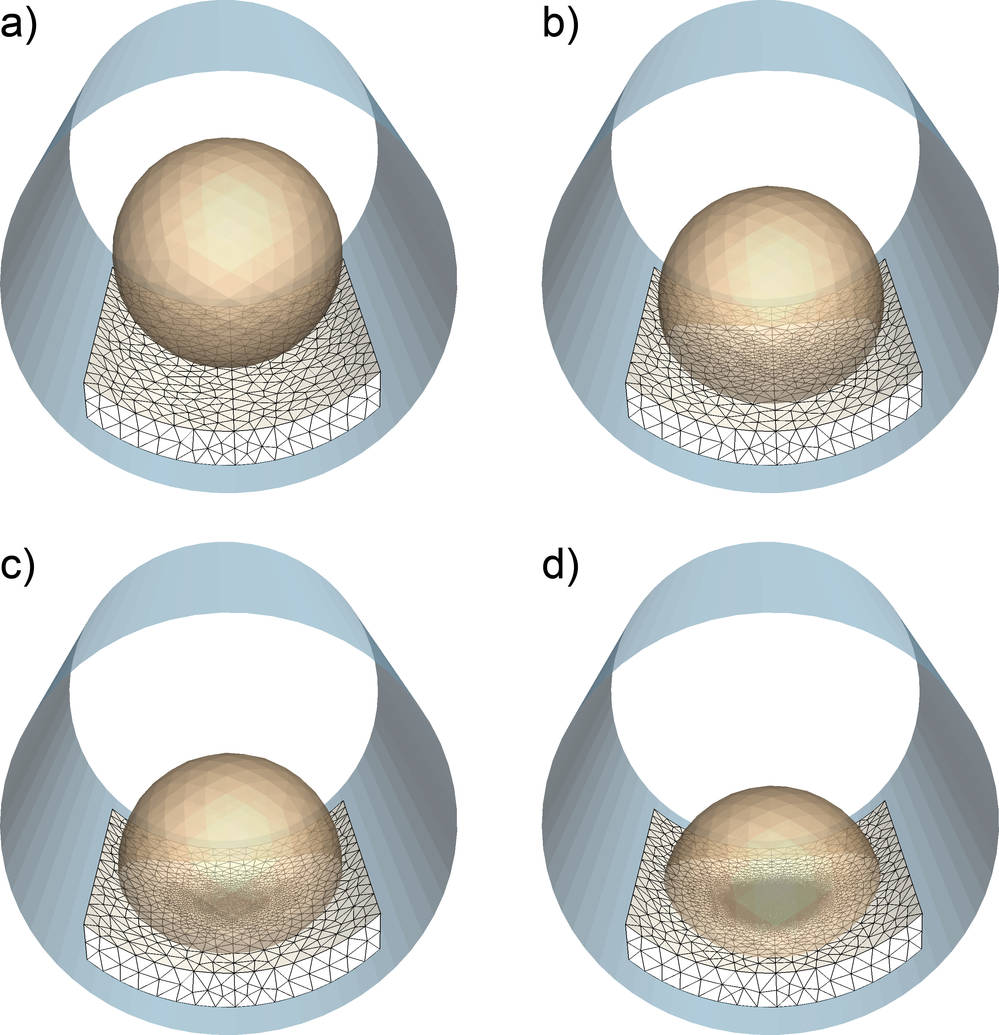}
\caption{Set of calotte structures acting as "refinement patches" for different distances of the MS to the boundary of the cylindrical sample.  As the MS reaches the surface, the transition is achieved with calotte meshes of appropriate discretization density, which are chosen dynamically from a predefined set of meshes according to the penetration depth of the MS into the surface. \label{Fig:calottes_example}}
\end{figure}
Regions of interest traced by a MS can move to the surface of the sample, so that the MS may leave partially the sample volume. In that case the MS must be cropped in order to preserve the shape and volume of the simulated sample, {\em i.e.}, nodes of the MS outside the volume are discarded. If the protruding part of the MS is removed, the remaining internal nodes and elements of the MS generally do not form a set that can serve as a replacement for the sample surface without further corrections. To maintain the surfaces of the sample mesh on one hand and to avoid a time-consuming remeshing of the MS on the other hand, a specific solution is needed, which in our case literally resembles a patch. If the MS is cropped, the gradual reduction of the discretization density is lost, resulting in a direct synchronization of nodes attached to very different cell size. To achieve a smooth transition, we use a set of typically 10 to 12 preprocessed micromagnetically treated calotte structures that connect the central parts of the sphere to the surface. A subset of typical calottes for a cylindrical structure is shown in Fig. \ref{Fig:calottes_example}. By using such "patches" we abandon the generality of the code, since the shape of each calotte is prepared to fit to the sample surface; in this case that of a cylinder. While it is straightforward to generalize this approach to other geometries, the preparation of the meshes must be performed for each specific geometry. The calottes used in our case  penetrate the sample by a thickness of twice the discretization size of the background mesh. The set of calottes is designed so that the change of discretization density is compatible to the MS at a given distance from the surface: If the MS barely touches the surface, there is no particular need for adjustments, but as the core of the MS approaches the surface, the calotte structure must be replaced by one that covers a large difference in discretization density. 

\subsection{Hierarchical coupling of the discretization grids\label{sec:coupling}}
So far, if we neglect the connecting boundary shells discussed before, we have four distinct discretization levels that enter into the simulation:
\begin{enumerate}
\item the atomistic Heisenberg structure in the core of the MS,
\item the micromagnetic outer part of the MS with radially increasing cell size,
\item the "patches" in the form of calotte structures with spatially varying discretization density, and 
\item the original mesh of the sample, which represents the background into which all the entities listed above are embedded.
\end{enumerate}
Each of these discretized regions is connected with a set of data for the magnetic structure, and they can partly or entirely share fractions of the volume of the sample. In order to obtain a consistent representation a hierarchical structure is defined according to the list shown above, so that the smallest available discretization cell is decisive. This means that the magnetic moments of the Heisenberg model are used to determine the magnetization direction at nodes of calotte structure located in the Heisenberg region, that the finer discretization in the calotte structure dictates the magnetization direction of the background mesh, etc. This is a simplified and general description in the sense that there is of course also, {\em e.g.}, a coupling of the background onto the MS; but the hierarchical order is in principle as listed above.
 
To couple the MS to the background, {\em i.e.}, to the ferromagnetic volume into which the Bloch point is embedded, each surface node of the MS receives the magnetic orientation and the effective field components by interpolation from the nodes of the tetrahedron of the static background mesh containing it. Conversely, a handshake between the regions is obtained as the data of each background node located inside the MS is assigned by the MS sphere, except for those background nodes which assign data to the boundary nodes of the MS themselves. 
In general, a transition region between the two meshes (outer MS sphere and background) could be included, similar to the one used to connect the Heisenberg  region with the micromagnetic inside the MS. In section \ref{sec:test} it is shown that better results are obtained without such an additional transition region.

\subsection{Intersections of multiscale spheres}
During a simulation, the inhomogeneities may distributed in space so that several multiscale regions and calotte structures are required simultaneously. The code can handle such situation, but some attention is necessary when different MS and/or calotte structures intersect. In order to synchronize the various regions we use a hierarchical approach to decide which mesh is chosen as the source of the magnetic properties (the donor) and which one is adapted to them (the acceptor). Several cases can be encountered:
\begin{enumerate}
\item If two MS are nearly concentric, the atomistic region of each sphere is overlapping but it is basically identical. There is no reason to favor one MS over the other in that purely atomistic region and there is no requirement to interpolate or to change anything.
\item When the atomistic region of one MS enters the continuum region of another MS, the values of the Heisenberg region are prioritized. The Heisenberg structure imprints the magnetization at any node of the finite element mesh in that region.
\item When two or more FEM regions of multiscale spheres overlap, the values at the nodes of the elements in one sphere may be adapted according to the values of the another sphere. For this, the role of the donor is first ascribed to the MS for which a given node is closest to the center point. This is however only a necessary, not a sufficient condition to discard the original values. To decide whether the values of the donor MS are used to interpolate onto the acceptor MS we introduce a further criterion, which is based on the position of the nodes of the tetrahedral element of the donor sphere containing the acceptor node.  We first determine the distance between these nodes and the center of the donor sphere and take the largest value. If this value is smaller than the distance between the acceptor node and the center of the acceptor sphere, the interpolation is performed.
Else the value remains unchanged, thereby preventing repeated interpolations that could introduce artifacts in the magnetic structure \cite{hertel_adaptive_1998}.
\item If a calotte structure is used, the magnetization is interpolated from the elements of the calotte acting as a donor onto the background mesh. If both, the donor and the acceptor mesh, are calotte structures an interpolation is performed only if the cell size of the donor cell is smaller than the smallest cell size in the Voronoi cell belonging to the acceptor node.
\end{enumerate}
In the code more details are considered in terms of donor and acceptor mesh, but most of these conventions are uncritical and the procedure described above should be sufficient to obtain a general picture of the complications arising and the procedures applied in the handshake regions of different meshes.


\subsection{Dynamics}
The dynamics of the magnetization within the multiscale model is governed by the Landau-Lifshitz-Gilbert equation. The equation is applied to calculate the orientation of the magnetic moments in the Heisenberg model and the direction of the  magnetization in the micromagnetic region -- both denoted here as $\vec{m}$.
\begin{equation}
 \frac{\dd \vec{m}}{\dd t} = - \gamma \vec{m} \times \vec{H}_{\text{eff}} + \alpha \left[ \vec{m} \times \frac{\vec{\dd m}}{\dd t} \right]  
\end{equation}
where $\gamma$ is the gyromagnetic ratio, $\alpha$ is the phenomenological Gilbert damping parameter. The effective field $\vec{H_{\rm eff}}$ is defined as:
\begin{eqnarray}
 \vec{H}_{\text{eff}}^{\rm MM} &=& -\frac{\partial e_{\rm tot}}{\mu_0M_s \partial \vec{m}} \\
 \vec{H}_{\text{eff}\ (i)}^{\rm Hei} &=& -\frac{V_i \partial e_{\rm tot}}{\mu_0\mu_i \partial \vec{m}} 
\end{eqnarray}
where $\mu_0$ is the vacuum permeability, $e_{\rm tot}$ the total free energy density, $M_s$ the saturation magnetization, $\mu_i$ the magnetic moment with index $i$ and $V_i$ its associated volume. 

We dynamically track the region of interest, {\em e.g.}, a Bloch point or (as discussed in the following test case) a vortex, and move the sphere if the region of interest is off-centered by a small number of lattice constants, usually one or two. Every propagation step moves the structure by an integer number of lattice vectors. This ensures that interpolations are performed only in the micromagnetic part, but not in the Heisenberg region of the MS. The position of the atoms remains fixed.

\section{\label{sec:test}Test of the code}

\begin{figure}
 \includegraphics[width=\linewidth]{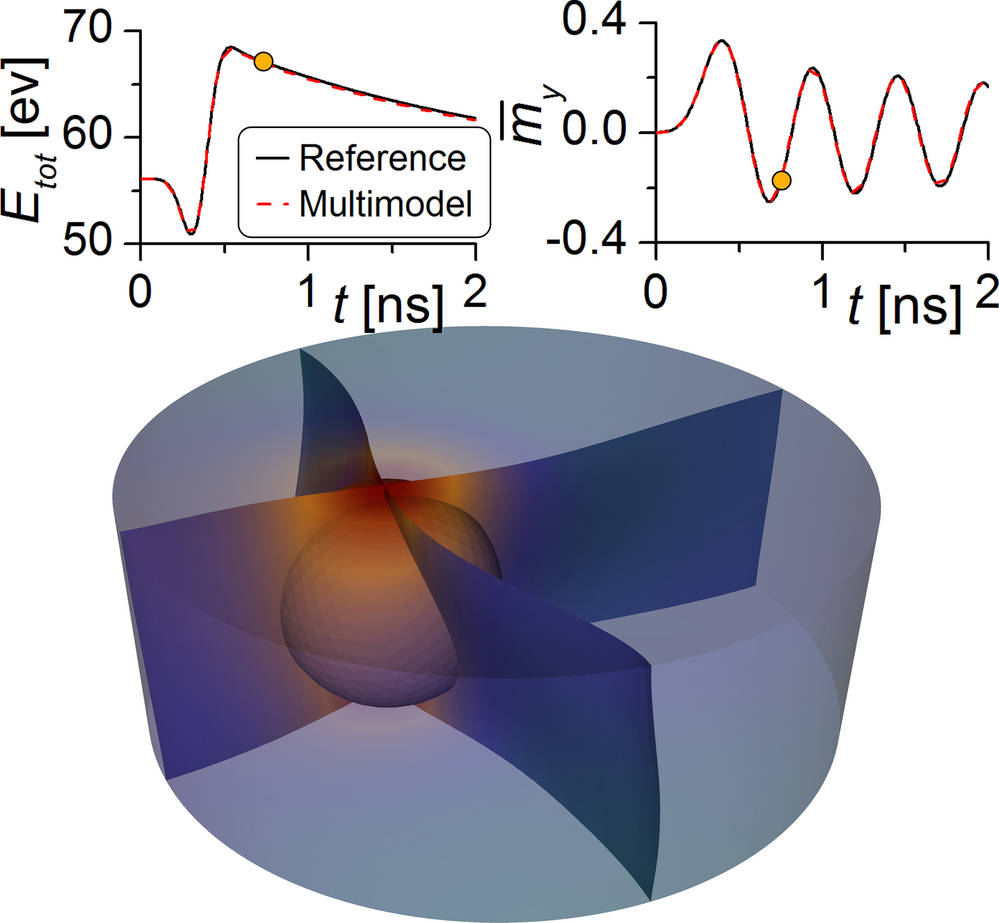}
\caption{\label{vtx-test}As a test case for the multimodel-multiscale code we use a Permalloy disc of 100~nm diameter and 35~nm  height with a vortex structure. A spatially homogeneous in-plane field pulse  with Gaussian time profile is applied to the relaxed vortex structure. The field pulse is applied along the $x$-axis; it has an amplitude of 50~mT, width $\sigma=100$~ps, and a peak-delay of $t_{\rm Max}=300$~ps. This perturbation of the system generates a well-known micromagnetic low-frequency excitation (gyrotropic motion) which represents a spiraling motion of the vortex core around the equilibrium position. The background mesh has a cell size of 1.7~nm. On the upper left, the evolution of the total energy of the system is displayed as a function of time, while on the upper right the spatially averaged and normalized $y$ component of the magnetization $\langle m_y\rangle$ is shown. The results obtained from the purely micromagnetic code and the multiscale-multimodel code are identical. The main frame on the bottom shows a snapshot of the magnetic structure at the moment indicated by the yellow dots in the upper frames. The color code represents the out-of-plane component $m_z$ of the magnetization and the semitransperent representation shows the position of the MS sphere at the core of the vortex. In the core of that sphere, the magnetic structure is calculated on an atomistic level. The crossing ribbons are the $m_x=0$ and $m_y=0$ isosurfaces, which we use routinely to track the position of the vortex (cf.~Ref.\cite{hertel_exchange_2006}). 
}
\end{figure}

Before studying the structure and the propagation of Bloch points with this complex algorithm, it is necessary to gain confidence into its capabilities and to test its reliability. Most importantly, it is necessary to ensure that the complicated nesting of meshes does not introduce numerical artifacts, like artificial oscillations, instabilities or increased damping. In order to investigate such effects, or the absence thereof, we study the gyrotropic motion of a vortex in a thin-film element. Such situations of gyrating vortices in patterned magnetic elements have been studied intensively over the past years and the physics is well understood (cf.~{\em e.g.} Ref.~\cite{choe_vortex_2004} and references therein). This example can be considered as a typical problem that modern micromagnetic simulation codes like {\tt TetraMag} can simulate with high accuracy; yet it is sufficiently non-trivial to unveil possible problems connected with the dynamic multiscale-multimodel code.

This case does not contain a singularity, hence an atomistic treatment is unnecessary. Nevertheless we introduce a MS here which resolves the core of the vortex with atomistic accuracy. The position of the vortex core can be traced conveniently by using the intersection of the isosurfaces $m_x=0$ and $m_y=0$ of the in-plane magnetization components (cf.~Ref.\cite{hertel_exchange_2006}). Since these isosurfaces can be tilted, so that their crossing line is not necessarily paralallel to the $z$ axis, we chose the intersection in the middle $z=t/2$ of the disk to determine the vortex position, where $t$ is the thickness and the bottom surface is at $z=0$.

The material parameters of the simulations are the typical values of Permalloy, with exchange stiffness $A=\SI{1.31e-11}{\joule/meter}$, saturation magnetization $\mu_0 M_s = \SI{1}{\tesla}$ and negligible magneto-crystalline anisotropy. The damping is set to $\alpha = 0.02$. 
The exchange length $l_{\rm exc}$ of this material is about \SI{5.7}{\nano\meter}, from which the vortex core radius can be estimated to \cite{usov_magnetization_1993} 
\begin{equation}
r_{\rm core} = 0.68 \ l_{\rm exc} \left( \frac{h}{l_{\rm exc}}\right)^{1/3} \approx \SI{6.1}{\nano\meter}
\end{equation}
In the vortex core, the magnetization rotates by \SI{180}{\degree}, which according to Eq.~\ref{Eq:errMM} corresponds to $\lambda\approx \SI{12.2}{\nano\meter}$. 
For the Heisenberg parameters we use a BCC lattice with lattice constant \nobreak{ $a=\SI{2.86}{\angstrom}$}, resulting in $\delta_G \approx \SI{2.2}{\angstrom}$. This yields a magnetic moment of $\mu = 1.01 \mu_B $ per lattice site. The Heisenberg exchange constant between nearest neighbors is \nobreak{$J = \SI{11.79}{\milli\electronvolt}$. By virtue of Eq.~\ref{Eq:errMM} we chose $R_1=\SI{7}{\nano\meter}$ with $\delta_T=\SI{2}{\nano\meter}$ in order to keep the local micromagnetic error below \SI{0.1}{\percent} for a \SI{180}{\degree} rotation within the diameter of the core region.  

As shown in Fig.~(\ref{vtx-test}) the dynamic multiscale-multimodel algorithm reproduces perfectly the results obtained by the micromagnetic code. The MS sphere follows the position of the vortex core, as can be seen in a movie provided as supplementary material\cite{supplement}. 
This agreement is reassuring because it demonstrates that in spite of the numerous interpolations, the simultaneous use of different models, and the enormously different length scales that are bridged at any time step of the calculation, the result remains unchanged. In the two frames shown on the upper part of Fig.~(\ref{vtx-test}) the evolution of the energy and the average magnetization component $\langle m_y\rangle$ are displayed as a function of time. In both cases, the data obtained from the multimodel-code matches perfectly the data of the micromagnetic code, so that the lines cannot be distinguished on that scale. In order to examine minor variations we look at this data in greater detail. More specifically, we use this data to investigate the necessity of an internal transition region between the MS sphere and the background. Such a transition region was previously introduced to connect the micromagnetic model with the Heisenberg calculation. In that case, the accuracy of the calculation increased with the width of the transition region. Remarkably, this is not true for the connection of the outer part of the MS sphere to the background. There, a direct connection of the nodes on the outer shell of the MS to the background gives the best results. This is shown in Fig.~(\ref{Fig:deviation_hei_tm}) where the previously described test case of a gyrating vortex is simulated with identical conditions but different thickness $\delta_T$ of a transition region at the boundary of the MS. This transition region is illustrated in Fig. \ref{Fig:implantedSphere} as the shading where the color changes from orange to green. While in all cases the deviations from the micromagnetic reference result are very small, the results clearly show that such a transition is detrimental to the accuracy. A direct connection of the nodes at the boundary of the MS sphere to the background leads to better agreement than a weighted averaging of the contributions from the outer regions of the MS and the background. The accuracy of the simulation is considerably lower when  $\delta_T$ exceeds about one nanometer. The results indicate that for the given discretization  it is best to discard any transition region between the two meshes and use the coupling rules described in \ref{sec:coupling} instead.

\begin{figure}
 \includegraphics[width=\linewidth]{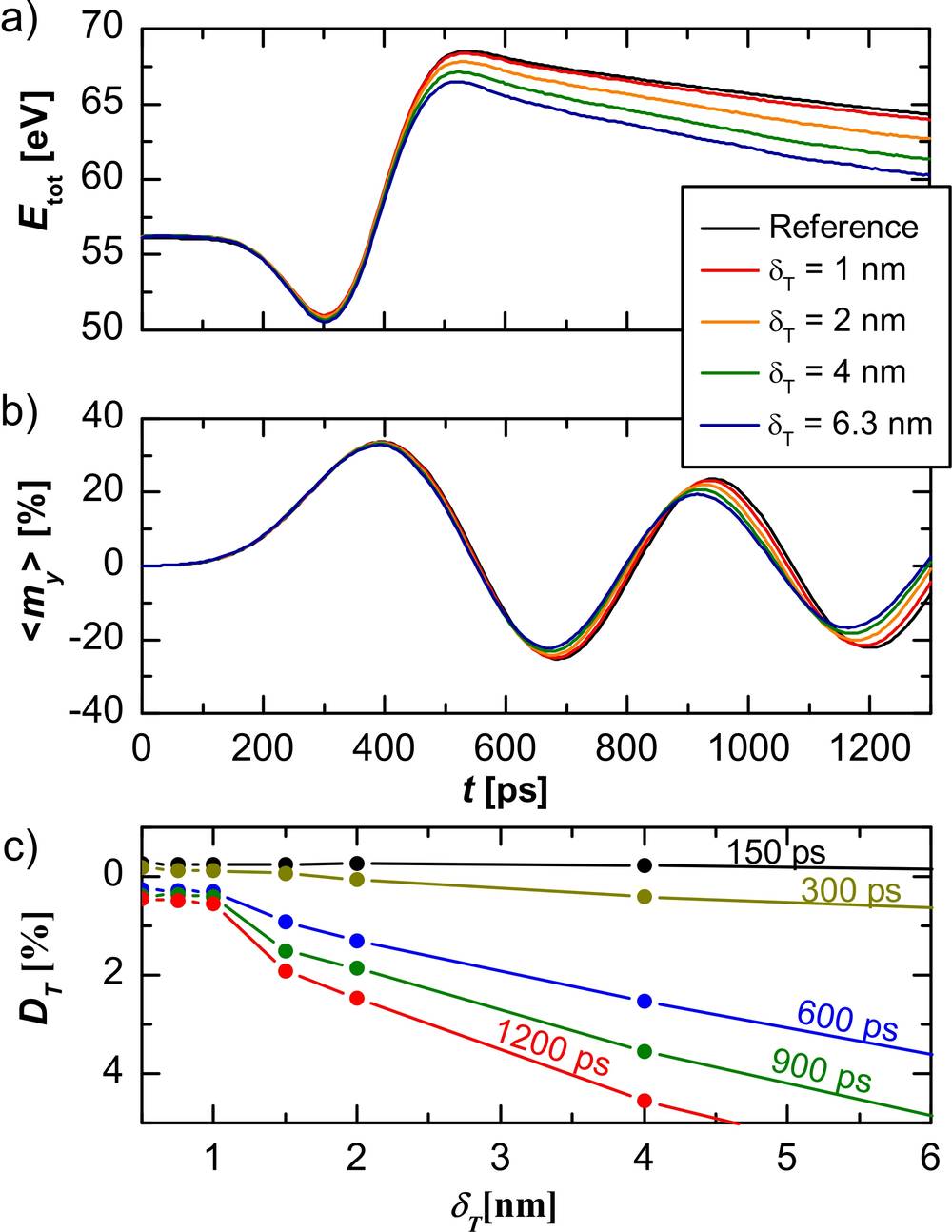}
\caption{Total energy (a) and average $y$-component of the magnetization  $m_y$ (b) for the previously discussed case of a gyrating vortex. The various lines correspond to the results obtained with transition regions of different thickness $\delta_T$. Panel (c) shows the error of the total energy connected with the multiscale-multimodel method as a function the thickness of the transition region $\delta_T$. Each of the lines refers to a moment in time (as indicated) after the beginning of the simulation at $t=0$. Here, the exact value of the energy is assumed to be the one obtained from the micromagnetic simulation, and the deviation from this value is displayed in percentage points\label{Fig:deviation_hei_tm}.}
\end{figure}

In addition to the simulations with nearest neighbor exchange interaction we studied the same systems with the Heisenberg exchange parameters of Fe as described by Pajda {\em et al.} \cite{pajda_ab_2001} with a total of 144 neighbors, scaled to match the micromagentic exchange stiffness $A=\SI{1.31e-11}{\joule/\meter}$ of Permalloy. This yielded no significant change of the results  compared to those obtained with a nearest-neighbor exchange interaction.

\section{\label{sec:blochpoint} Bloch point motion in a soft-magnetic cylinder}

\begin{figure}
\includegraphics[width=\linewidth]{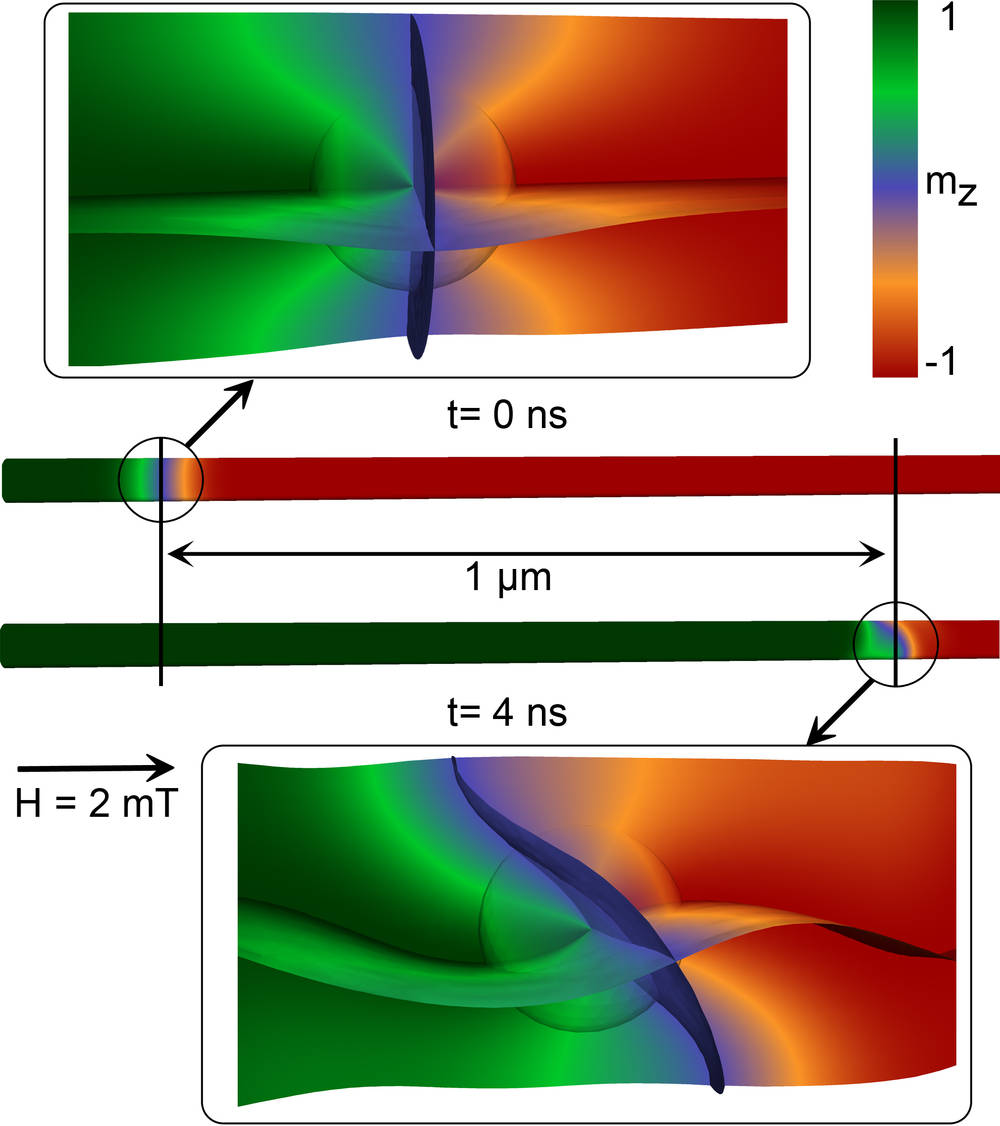}
\caption{Snapshots of the simulated vortex domain wall (DW) propagation in a ferromagnetic cylinder \cite{ref-mat} with a Bloch point in its core. The two magnified images show the magnetic configuration around the Bloch point in the equilibrium configuration and after applying an external field of \SI{2}{\milli\tesla} for four nanoseconds. The crossing of the three $m_{\{x,y,z\}} = 0$ isosurfaces indicates the position of the BP in the center of the multimodel sphere, and the coloring  represents the axial magnetization component. \label{fig:BPpropH2mT}}
\end{figure}

An example for the propagation of Bloch points is the magnetization reversal process of a soft magnetic cylinder of sufficiently large diameter, where head-to-head or tail-to-tail domain walls develop as vortex domain walls hosting a Bloch point in their center. As the domain walls are moved, {\em e.g.}, by means of an external magnetic field, the Bloch point propagates along the cylinder axis. Earlier publications \cite{arrott_point_1979,thiaville_micromagnetic_2003,hertel_magnetic_2004} studied this type of reversal process in the framework of pure micromagnetism, even though the Bloch point requires a multimodel investigation method. Using our multimodel framework we examined the propagation of a Bloch point on a \SI{4}{\micro\meter} long cylinder with a diameter of \SI{60}{\nano\meter}, with material and simulation parameters according to Ref.~\cite{ref-mat}. 

In the equilibrium configuration, the Bloch point resides in the center of the vortex domain wall and the configuration displays cylindrical symmetry. For very low external fields the Bloch point experiences a pinning, as predicted by theory\cite{Reinhardt1973}. A detailed analysis of the Bloch point pinning at atomic lattice sites will be the subject of a forthcoming paper. If a sufficiently large field is applied to overcome this pinning, the Bloch point and the domain wall propagates smoothly in axial ($z$) direction. Figure \ref{fig:BPpropH2mT} shows two snapshots of the Bloch point propagation for an applied axial field of $H_{\text{ext}} = \SI{2}{\milli\tesla}$. During a short acceleration period, the cylinder symmetry is preserved, but the magnetic configuration changes when the domain wall reaches its final velocity of approximately $v \approx \SI{300}{\meter/\second}$. Before reaching this final velocity, the $m_z=0$ isosurface tilts and eventually encloses an angle of \SI{45}{\degree} between its normal vector and the  $z$-axis. This corresponds to a distortion of the magnetic structure around the Bloch point, including an unforeseen spontaneous break of symmetry. Nevertheless, the Bloch point remains centered and does not lag behind the domain wall.


\section{\label{sec:Concl}Summary}
We have presented in detail the implementation of a dynamic multiscale-multimodel algorithm which combines a micromagnetic finite element code with a classical Heisenberg model. The code runs entirely on graphic cards, which allows for a fast computation of the complex interacting meshes involved in the simulation. An analytical estimation has been derived to calculate the critical distance between misaligned magnetic moments below which an atomistic treatment becomes decisive. This critical size is used as a criterion for the minimum diameter of the region in the model in which exchange interaction is treated purely by the Heisenberg model. We have shown that the numerically critical part of the method, {\em i.e.}, the interface between sample and the implanted multiscale sphere, gives rise to a local error of less then one percent; which is good enough to use the method for reliable future studies on the dynamics of Bloch points. The code has been thoroughly tested, ensuring that the simultaneous treatment of different methods does not introduce artifacts. The multiscale region consist of a preprocessed sphere that can be moved through the sample without the need of numerically expansive remeshing processes. Owing to this ability of tracing regions of interest dynamically, we could demonstrate an example study on the dynamics of Bloch points with unprecedented accuracy. 

The dynamics of Bloch points can be expected to contain a number of surprising effects, similar to the rich variety of effects that have been discovered concerning the dynamics of vortices and domain walls. The algorithm presented in this article is the first of this kind, which enables a precise study of the Bloch point dynamics. More studies on this topic will be published soon.


\bibliography{MS-code_final-refs}

\begin{thebibliography}{31}
\bibitem [{ref()}]{ref-mat}%
  \BibitemOpen
  \href@noop {} {}\bibinfo {note} {The material of the 4~$\mu$m long cylinder
  with 60~nm diameter is a bcc crystal with lattice constant $a=2.866$\AA, a
  magnetic moment of one Bohr magneton per lattice site. The atomistic model
  considers nearest-neighbor interaction with an exchange constant of
  $J=11.79$~meV. In micromagnetic terms, these paremeters correspond to a
  saturation magnetization of $\mu_0 M_s= 1.00$~T and an exchange stiffness
  $A=1.31\times10^{-11}$ J/m. The sample (background) mesh consists of
  tetrahedrons with 3~nm cell size, resulting in 494914 vertices. The
  multimodel sphere has a total diameter of 37~nm, an atomistic core region
  with a diameter of 14~nm, including a model transition region of 2~nm
  width.}\BibitemShut {Stop}%
\end{thebibliography}
\include{MS-code_final-refs.bbl}

\end{document}